\documentclass[twocolumn,onecolappendix]{aastex631}

\newcommand{\rhigh}{R_{\rm high}}

\newcommand{\spin}{a_{\ast}}
\newcommand{\PA}{\mathrm{PA}}
\newcommand{\uv}{(u,v)}
\newcommand{\omegava}{\Omega_{\mathrm{VA}}}

\usepackage{hyperref}
\usepackage{amsmath}
\usepackage{placeins} 

\shorttitle{EHT Pattern Speeds in the Visibility Domain}
\shortauthors{Conroy et al.}

\graphicspath{{./}}

\begin{document}

\title{Event Horizon Telescope Pattern Speeds in the Visibility Domain}

\author[0000-0003-2886-2377]{Nicholas S. Conroy}
\affiliation{Department of Astronomy, University of Illinois at Urbana-Champaign, 1002 West Green Street, Urbana, IL 61801, USA}

\author[0000-0002-5518-2812]{Michi Baub\"ock}
\affiliation{Department of Physics, University of Illinois at Urbana-Champaign, 1110 West Green Street, Urbana, IL 61801, USA}
\affiliation{Illinois Center for the Advanced Study of the Universe, University of Illinois at Urbana-Champaign, 1110 West Green St., Urbana, IL 61801, USA}

\author[0000-0001-6765-877X]{Vedant Dhruv}
\affiliation{Department of Physics, University of Illinois at Urbana-Champaign, 1110 West Green Street, Urbana, IL 61801, USA}
\affiliation{Illinois Center for the Advanced Study of the Universe, University of Illinois at Urbana-Champaign, 1110 West Green St., Urbana, IL 61801, USA}

\author[0000-0002-3350-5588]{Daeyoung Lee}
\affiliation{Department of Physics, University of Illinois at Urbana-Champaign, 1110 West Green Street, Urbana, IL 61801, USA}

\author[0000-0001-6337-6126]{Chi-kwan Chan}
\affiliation{Steward Observatory and Department of Astronomy, University of Arizona, 933 N. Cherry Ave., Tucson, AZ 85721, USA}
\affiliation{Data Science Institute, University of Arizona, 1230 N. Cherry Ave., Tucson, AZ 85721, USA}
\affiliation{Program in Applied Mathematics, University of Arizona, 617 N. Santa Rita, Tucson, AZ 85721, USA}

\author[0000-0002-2514-5965]{Abhishek V. Joshi}
\affiliation{Department of Physics, University of Illinois at Urbana-Champaign, 1110 West Green Street, Urbana, IL 61801, USA}
\affiliation{Illinois Center for the Advanced Study of the Universe, University of Illinois at Urbana-Champaign, 1110 West Green St., Urbana, IL 61801, USA}

\author[0000-0002-0393-7734]{Cora Prather}
\affiliation{Black Hole Initiative at Harvard University, 20 Garden Street, Cambridge, MA 02138, USA}

\author[0000-0001-7451-8935]{Charles F. Gammie}
\affiliation{Department of Astronomy, University of Illinois at Urbana-Champaign, 1002 West Green Street, Urbana, IL 61801, USA}
\affiliation{Department of Physics, University of Illinois at Urbana-Champaign, 1110 West Green Street, Urbana, IL 61801, USA}
\affiliation{NCSA, University of Illinois at Urbana-Champaign, 1205 W. Clark St., Urbana, IL 61801, USA}
\affiliation{Illinois Center for the Advanced Study of the Universe, University of Illinois at Urbana-Champaign, 1110 West Green St., Urbana, IL 61801, USA}

\begin{abstract}

The Event Horizon Telescope (EHT) is preparing to produce time sequences of black hole images, or videos. 
In anticipation, we previously developed an autocorrelation technique to measure apparent rotational motion using the image-domain pattern speed $\Omega_p$.  Here, we extend this technique to the visibility domain and introduce the visibility amplitude pattern speed $\omegava$.  We show that in the Illinois v3 library of EHT source models, $\omegava$ depends on the source inclination, black hole mass, black hole spin, accretion state (MAD or SANE), and baseline length, and then provide approximate fits for this dependence.  We show that $\omegava$ is particularly sensitive to baseline length for MAD (strongly magnetized) models, and that the slope of this dependence can be used to constrain black hole spin.  As with $\Omega_p$, models predict that $\omegava$ is well below the Keplerian frequency in the emission region for all model parameters. This is consistent with the idea that $\omegava$  measures an angular phase speed for waves propagating through the emission region. Finally, we identify the information that would be provided by space-based millimeter very long baseline interferometry such as the proposed BHEX mission.  
\end{abstract}
\keywords{Black holes (162), Magnetohydrodynamical simulations (1966), Very long baseline interferometry (1769), Accretion (14), Astrophysical fluid dynamics (101)}

\section{Introduction}\label{sec:intro}

The Event Horizon Telescope (EHT) has captured the first horizon-scale images of the black holes M87* and Sgr A* \citep{M87PaperI, SgrAPaperI}. The EHT is now capable of producing time sequences of images, or videos. These videos are made possible through the regular revisiting of sources, improved dynamical imaging algorithms, and continually improving baseline coverage for the EHT telescope array \citep{Doeleman_2023_ngeht, Johnson_2023_ngeht, Midrange_science_goals, Ayzenberg_2025_ngEHT}. 

One key feature expected in EHT videos is {\em rotation}.  In earlier work \citep{Conroy_2023}, we developed a strategy for measuring rotation in the image domain.  The method assumed that the images were ring-like and that a spectrum of fluctuations in brightness on the ring were resolved and well sampled over time.  By autocorrelating fluctuations over position angle ($\PA$) and time, we measured a mean angular speed for the fluctuation spectrum, which we referred to as the pattern speed $\Omega_p$. 

\citet{Conroy_2023}  surveyed $\Omega_p$ in a library of models for EHT sources.  The models consist of general relativistic magnetohydrodynamic (GRMHD) simulations that have been imaged using a relativistic ray-tracing scheme \citep{Wong_2022, Dhruv_grmhd_survey_2024}.  These libraries have been largely successful at describing the EHT sources M87* and Sgr A*, although the majority of models in the library are excluded by the data. There are best-bet models, however, that pass all $10/10 = 100\%$  observational constraints for M87* \citep[see Tables 2, 2–3, and 3 in][respectively]{M87PaperV, M87PaperVIII, M87PaperIX}, and $18/19 \approx 95\%$ of observational constraints for Sgr A* \cite[Tables 4–5 and 1 in][respectively]{SgrAPaperV, SgrAPaperVIII}. Importantly, the GRMHD models provide predictions for the dynamics in EHT videos.

A natural expectation for rotation in EHT videos is based on ``hot-spot'' models, in which a luminous region orbits with the inflowing plasma.  In X-ray astronomy, this idea can be traced to \citet{Sunyaev_1972}.  For EHT sources, there are numerous applications of hot-spot models, beginning with \citet{Broderick_loeb_hotspot1, Broderick_loeb_hotspot2, Broderick_loeb_hotspot3}, who made predictions for EHT observations of Sgr A* based on a hot-spot model. More recently, theoretical studies of hot-spot videos have used correlation analysis similar to \cite{Conroy_2023} to extract signals of lensing (\citealp{Wong_2021_glimmer, Zhang_2025, Zhu_2025}). Hot-spot modeling was also applied to interpret fluctuations seen in near-infrared astrometry and polarization by the GRAVITY Collaboration and in millimeter polarization by the Atacama Large Millimeter/submillimeter Array \citep{GRAVITY_2018, GRAVITY_2020_flares, Wielgus_22_hotspot,Yfantis_24_hotspot, Yfantis_24_hotspot_2, Ricarte_2025}. Unlike in \citet{Conroy_2023}, here we examine the apparent rotational velocity of features in the image and visibility domain.

For a hot spot on an equatorial circular orbit just subject to gravitation at a radius $r$ from the black hole, the corresponding Keplerian velocity is $\Omega_K = 57.3 (r + a_*)^{-3/2}$ deg per $G Mc^{-3}$ (hereafter deg $t_g^{-1}$ for $t_g\equiv G Mc^{-3}$), where $a_*$ is the dimensionless black hole spin.  GRMHD models suggest that in EHT sources, emission is dominated by plasma at $r \approx 4 \, GMc^{-2}$. For $\spin = 0$, $\Omega_K(r = 4 \, GMc^{-2}) = 7.2$ deg $t_g^{-1}$. One of the main predictions of \citet{Conroy_2023}, based on a survey of GRMHD models, is that $\Omega_p$ is sub-Keplerian with a mean magnitude $\sim 1$ deg $t_g^{-1}$, i.e., that $\Omega_p \ll \Omega_K$.  

The emitting plasma need not be moving on circular orbits.  The orbital frequency of plasma in a subset of models (strongly magnetized, or MAD, models) is less than $\Omega_K$ by a factor of $\sim2$ at $r \sim 4 \, GMc^{-2}$ \citep{Dhruv_grmhd_survey_2024}.  This is still large compared to $\Omega_p$ in synthetic videos, suggesting that $\Omega_p$ measures an angular phase velocity of waves rather than entropy fluctuations that co-orbit with the inflowing plasma.  Indeed, Section 3.1 of \citet{Conroy_2023} showed that $\Omega_p$ measured in synthetic images closely matches the phase speed of pressure fluctuations in the underlying GRMHD simulation.  

Setting aside the {\em magnitude} of $\Omega_p$, it is interesting even to measure its {\em sign}, since models predict this reveals the direction of rotation of the inflowing material.  Knowing the sense of rotation would vastly reduce the model space, especially for M87* where the ring asymmetry indicates the black hole spin vector is pointed away from Earth \citep{M87PaperV, Bernshteyn_2026, Conroy_2026_PA}.  

Why is knowing the sense of rotation interesting? It is possible that the accretion flow is retrograde, i.e., the black hole spin vector and accretion angular momentum vector are antialigned. For M87*, a movie would enable this to be shown convincingly, with ring asymmetry providing the black hole spin direction and $\Omega_p$ providing the accretion flow orbital angular momentum.  Then, if the jet is shown to rotate with the hole but not the accretion flow, there could be only one conclusion: that frame dragging of field lines was driving the jet.  This would be a powerful test of the \citet{Blandford_Znajek} model (see Appendix \ref{subsec:v5}).

However interesting it may be to measure $\Omega_p$, the procedure introduced in \citet{Conroy_2023} requires reconstruction of a time-sequence of images.  Since images are a nonunique representation of very long baseline interferometry (VLBI) data (EHT sparsely samples the Fourier transform of the image), one might be concerned that regularization procedures used in image reconstruction could introduce spurious correlations and therefore spurious $\Omega_p$ measurements.  Here, we take a step toward measuring the apparent rotation rate in a reconstruction-independent fashion.

In this paper, we introduce a pattern speed $\omegava$ that is measured from visibility amplitudes in the $\uv$ domain.  The methodology is similar to \citet{Conroy_2023}.  
We measure $\omegava$ assuming complete $\uv$ coverage and no observational noise. We leave the problem of extracting $\omegava$ from more realistic, noisy, inconsistently sampled visibility amplitudes to future work. 
Section \ref{sec:method} reviews the methodology and changes necessary to apply these methods in the $\uv$ domain. In Section \ref{sec:results}, we survey a library of GRMHD models and present the results.  As in the image domain, visibility domain pattern speeds vary with radius (here, baseline length) and model parameters, and can thus potentially be used to constrain black hole mass, inclination, and spin. We extend our analysis to space VLBI in Section \ref{sec:space}, and conclude in Section \ref{sec:conclusion}.   

\section{Expectations for the Visibility Domain}
\label{sec:expect}

\begin{figure*}
   \plotone{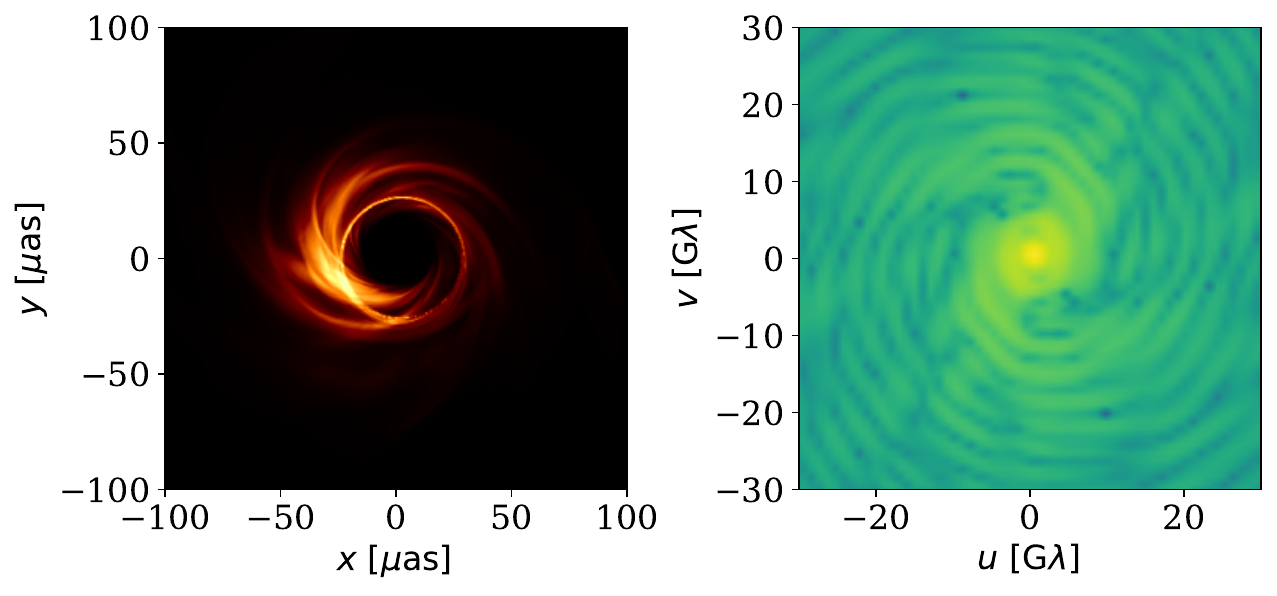}
   \caption{A 230 GHz snapshot image from a fiducial Sgr A* model (MAD, $\spin = 0.5$, $i = 30^\circ$, $\rhigh = 160$) (left), and the accompanying visibility amplitudes (right). The visibility amplitude color map is logarithmic. The image-domain ring produces a series of damping lobes and nulls, while spiral waves produce spiral features in the $\uv$ domain. The helicity of the spiral has flipped, as expected.
   }
   \label{fig:frame} 
\end{figure*}

The dominant feature in EHT images of M87* and Sgr A* is a ring. The ring coincides approximately with the critical curve \citep{Gralla_2019}, which separates geodesics that intersect the horizon from those that do not; peak brightness occurs near the critical curve  because synchrotron emission drops off steeply with radius in the accretion flow.  

VLBI experiments like EHT measure {\em visibility amplitudes}, which are the Fourier amplitudes of the images, and {\em closure phases}, which are combinations of Fourier component phases.  The Fourier phases are not directly measurable due to atmospheric phase fluctuations; closure phases combine the phases on individual baselines (i.e., for individual Fourier components) so that atmospheric phase fluctuations at individual antennas cancel out \citep[e.g.,][]{Thompson_2017, Blackburn_2020}.  We leave the extraction of rotation information from closure phases to future work; here we focus exclusively on visibility amplitudes.

A narrow ring with angular structure becomes a sum of Bessel functions in the $\uv$ domain \citep{Johnson2020_photonring}. Extended structure adds complexity to the image.  Visibility amplitudes for GRMHD models have a central peak at $\rho^2 = u^2 + v^2 = 0$; $\rho$ also corresponds to projected baseline length measured in wavelengths in the VLBI array.  The visibility amplitudes then have a series of nulls and lobes of gradually decreasing amplitude. The first null lies at $\rho \approx 0.38/b_0$, where $b_0 = \sqrt{27} \, GMc^{-2}D^{-1}$ is the ring's angular radius, or impact parameter, measured in radians.  In 2017, EHT coverage approximately spanned $\rho \in [0, 9) \, G\lambda$ for Sgr A*. This is enough to cover two lobes and two nulls, as well as the start of the third lobe \citep{SgrAPaperI}. 

An example image-domain frame from a typical Sgr A* model and the corresponding $\uv$ visibility amplitudes are shown in Figure \ref{fig:frame}. 
In addition to a ring and extended structure, the image contains superposed narrow  spiral features.  These features can be modeled as segments of a  logarithmic spiral.  In the image domain, a logarithmic spiral is $\propto e^{i(\alpha \ln r \, + m\theta)}$ for polar coordinates $(r, \theta)$, cotangent of the pitch angle $\alpha/m$, and azimuthal mode number $m$.  Lines of constant phase lie at $\alpha \ln r + m \theta = const.$  Using polar coordinates $\rho,\phi$ in the Fourier domain, the Fourier transform of the logarithmic spiral is $\propto e^{i(-\alpha \ln \rho \, + m\phi)}$ for which lines of constant phase lie at $-\alpha \ln \rho + m \phi = const$. That is, spiral features in the image domain map into spiral features in the $\uv$ domain, with the sense of the spiral reversed. This is consistent with the inversion in the sense of the spiral shown going from the left to right panel in  Figure \ref{fig:frame}.

A source that can be approximated by a Gaussian of standard deviation $\sigma$ in the image domain will have visibility amplitudes that are coherent over regions of half-width $\Delta \rho \sim (2\pi \sigma)^{-1}$ in the $\uv$ domain. For Sgr A*, $\Delta \rho \sim 1.6 
\mathrm{G}\lambda$ (assuming FWHM = $47 \,\mu$as, so the second moment of the fitted Gaussian is $20\,\mu$as).  Thus, the $\uv$ spirals cohere over a region of $\Delta \rho \sim 1.6 \, \mathrm{G}\lambda$. 

If the spiral circulates around the ring with time and $\PA$ dependence $\propto e^{i m\theta - i \omega t}$, then the $\uv$ time dependence will be the same, and in both cases, the spiral (or any other feature with sinusoidal position angle dependence) will have angular phase speed $\omega/m = \Omega_p=\omegava$. That is, the angular speed of features is the same in the $\uv$ domain and the image domain.  

We do not expect, however, to observe a single phase speed for all features in the image: the image contains a spectrum of fluctuations.  Then $\Omega_p$ measures the dominant phase speed, which corresponds to the peak of the spectrum of angular phase speeds. 

The pattern speed signal is strongest on the ring where emission is brightest.  There is also a signal at larger impact parameter (distance from the center of the ring).  GRMHD models exhibit a pattern speed ``rotation curve,'' where $\Omega_p$ depends on impact parameter, defined as distance from the center of the ring. The models exhibit differential rotation, with slower pattern speed at larger radii and faster pattern speed closer to the ring. The slope of this rotation curve varies with spin. This suggests that there may be a pseudo-rotation curve for pattern speeds measured in the $\uv$ domain, as a function of baseline length $\rho$. In the following section, we describe how to measure this $\uv$ pseudo-rotation curve. We assume complete $\uv$ coverage to map trends as a function of model parameters, and leave the task of accurate measurement in realistic observational conditions to future work.  

\section{Measurement Technique}\label{sec:method}

Our technique is based on the methodology for measuring image-domain pattern speeds described in \citet{Conroy_2023}.  We refer the reader to that publication for full details. The methodology, along with the minor refinements necessary to adapt it to the $\uv$ domain, are summarized here in Sections \ref{sec:cylinder} and \ref{sec:method_pattern}. 

\subsection{Model Library}

\label{sec:models}
The analysis is performed on synthetic data from the Illinois Sgr A* ``v3'' model library \citep{Dhruv_grmhd_survey_2024}. The GRMHD library was run with KHARMA \citep{Prather_kharma_2024}, which is based on \citet{Gammie_03}. It was imaged with {\tt ipole} \citep{Mosc_2018}. 

The library consists of 360 models, each lasting $15000 \, t_g \approx 85$ hr with frames every $5 \, t_g \approx 100$ s, spanning four parameters. The parameters are the dimensionless black hole spin  $\spin \equiv Jc/GM^2$, the magnetic flux (strongly magnetized ``MADs'' or weakly magnetized ``SANEs''), the inclination angle $i$ (where $i = 0$ indicates the angular momentum vector of the disk is pointed toward the observer), and the electron temperature parameter $\rhigh$ (the maximum ratio of the ion-to-electron temperature).

\citet{SgrAPaperV} and \citet{Wong_2022} provide additional details for the model library.  In \citet{Conroy_2023} and below (Section \ref{sec:conclusion}), we discuss the impact of model limitations on the pattern speed.

\subsection{Cylinder Plots} 
\label{sec:cylinder}

The synthetic visibility amplitudes are the amplitudes $A=A(u,v)$ of Fourier transformed images from the model library.  As discussed above, we expect features to fluctuate and rotate in the $\uv$ domain.  

\begin{figure*}
   \plotone{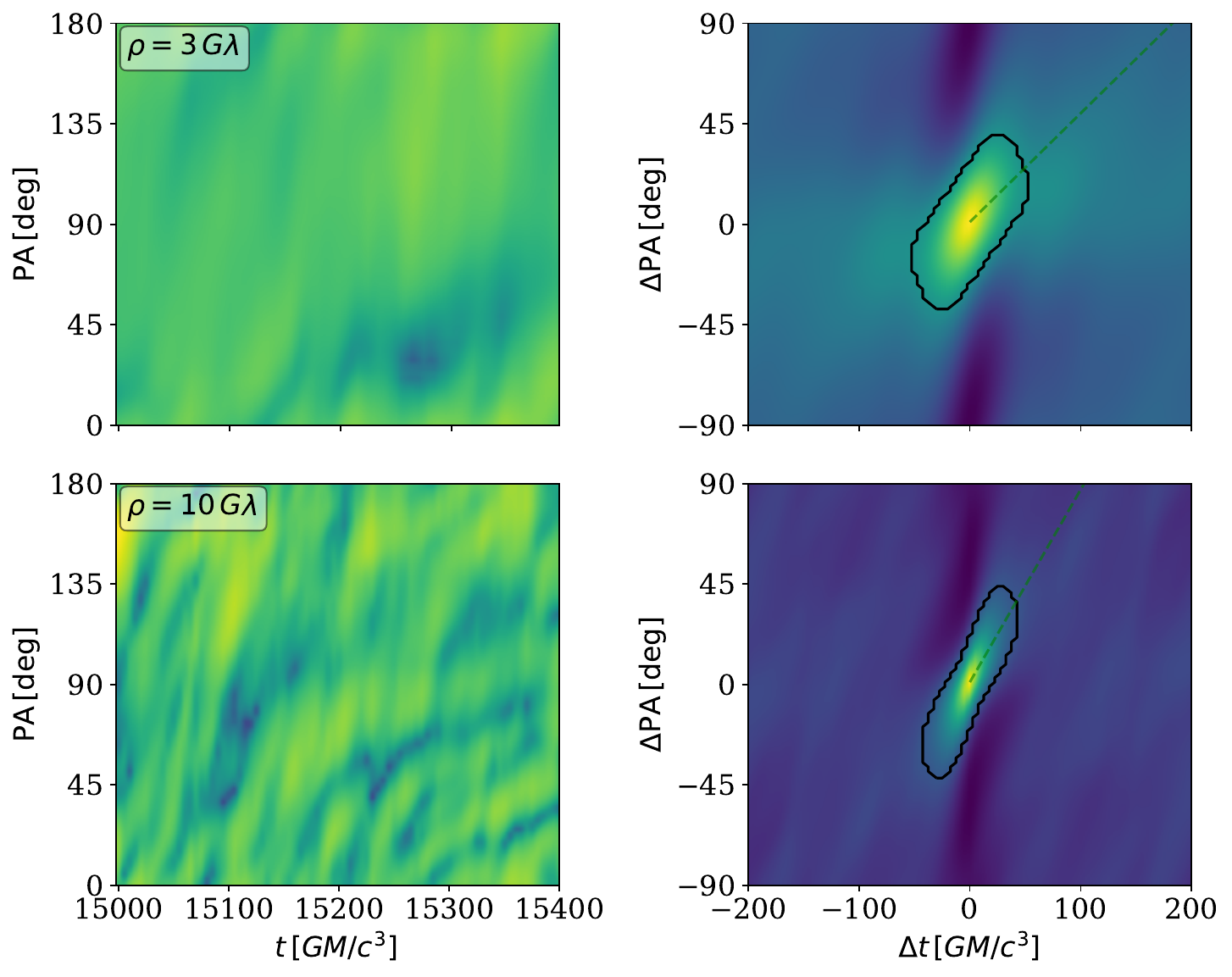}
   \caption{A $400 \, t_g$ window of a $\uv$ cylinder plots (left) and the autocorrelations (right) from the fiducial Sgr A* model (MAD, $\spin = 0.5$, $i = 30^\circ$, $\rhigh = 160$), measured at $\rho = 3 \, G\lambda$ (top) and $10 \, G\lambda$ (bottom). The bounded region of the autocorrelation $\xi > \xi_{\rm crit}$ is surrounded by a black contour, and the measured pattern speed $\omegava$ is marked by the dashed green line. Notice the size and slope of the central correlated peak changes with varying $\rho$.
   }       
   \label{fig:cylinder} 
\end{figure*} 

We perform a coordinate transformation on the visibilities. 
We sample the visibility amplitudes $A(u,v)$ along a ring in polar coordinate $\uv=(\rho \cos \PA, \, \rho \sin \PA)$, for radius $\rho$ and position angle $\PA$, across every frame at time $t$. Thus, we sample $A=A(\PA, t; \rho)$.
This forms a ``cylinder plot'' at each $\rho$. The motion of bright visibility features at constant $\rho$ will appear in the cylinder plot as diagonal bands, with slope corresponding a change in $\PA$ over time.  This slope corresponds to their apparent angular velocity. Since visibility domain structure and EHT coverage spans numerous $\uv$ radii,  we produce cylinder plots at multiple $\rho$ to form a pseudo-rotation curve. Example cylinder plots for one model at $\rho = 3$ and $10 \,G\lambda$ are shown on the left half of Figure \ref{fig:cylinder}.  

Unlike the image domain, the $\uv$ domain is  symmetrical about $\Delta \PA \pm 180^\circ$, so we crop the $\uv$ cylinder plot to span only $\Delta \PA = 180^\circ$, corresponding to a semicircle in the $\uv$ domain. The orientation of the  semicircle does not matter.

As in the image domain, we normalize the cylinder plot to produce $\tilde{A}(\PA, t; \rho)$. We find $\tilde{A}$ by taking $\log(A)$ then mean subtracting along every $\PA$ and $t$ slice. 
Normalization corrects for mean brightness asymmetries in the visibility domain, such as those caused by the large-scale brightness asymmetry in $\PA$ \citep{Bernshteyn_2026} and brightness variations in $t$, which would otherwise dominate the measurement. The final $\uv$ cylinder plot is thus a measure of relative fluctuations in the visibility amplitudes $A$ along a ring over time: \begin{equation}
\begin{split}
    \tilde{A}(\PA, t; \rho)  \equiv \ln A (\PA, t) \\
    - \langle \ln A\rangle_{\PA} - \langle \ln A\rangle_{t}
\end{split}
\end{equation}

\subsection{Pattern Speeds in Visibility Amplitudes} 
\label{sec:method_pattern}

To estimate the visibility amplitude pattern speed, we calculate the autocorrelation $\xi(\Delta t, \Delta \PA)$ of $\tilde{A}$. The autocorrelation has a central peak at  $(\Delta t, \Delta \PA) = (0,0)$, and we identify the visibility amplitude pattern speed $\omegava$ as the slope around the central peak. The sign of the slope indicates the direction of rotation (clockwise or counterclockwise about the ring). Two example $\xi$s are shown in Figure \ref{fig:cylinder}. Notice that a positive slope indicates that visibility features tend to move counterclockwise on the sky, toward larger $\PA$. 

We use a second moment method to estimate $\omegava$, setting
\begin{equation}
\omegava \equiv (M_{\Delta\PA\,\Delta t}/M_{\Delta t \, \Delta t} ) \, |_{\xi \in \mathcal{R}}
\label{eq_omegap}
\end{equation}
Here, the second moments are $M_{xy} \equiv   \int x y \, \xi \, d{\Delta\PA} \,d\Delta t$. Plugging in, we see the units are deg $t_g^{-1}$, as expected. The integration is defined within the bound region $\xi \in \mathcal{R}$. The boundary is optimized to enclose the peak in the autocorrelation, where the signal for rotating correlated features is strongest. In most cases, the integration region corresponds to $\xi>\xi_{\rm crit}$. However, we also exclude rare spurious peaks or uncorrelated noise that surpass the threshold $\xi_{\rm crit}$ but are detached from the central correlated peak (see the end of this section). A derivation of Equation \ref{eq_omegap} in the image domain can be found in \citet{Conroy_2023}.

For the image-domain analysis, we set $\xi_{\rm crit} = 0.8$ after optimizing the threshold in test problems \citep{Conroy_2023}.  In the $\uv$ domain analysis, the situation is more complicated.  As $\rho$ varies from $\approx 1 \, G\lambda$ to $> 10 \, G\lambda$, the correlation time changes from $\tau > 100 \, t_g$ to $\tau \approx 10 \,  t_g$ (see Figure \ref{fig:cylinder}). The size and shape of the correlated region changes with changing $\rho$, so a fixed $\xi_{\rm crit}$ is unlikely to be optimal.  

We have therefore experimented with varying the integration bounds in a way that depends on the signal to noise ratio in $\xi$. The distribution of noise in $\xi$ (which is largely due to finite sample size in time) is approximately normal with mean $\mu = 0$; the signal is in the positive tail of the distribution.  We fix $\xi_{\rm crit} = 3\sigma$, where the threshold is set empirically to optimize accuracy in test problems (see Section \ref{sec:tests}). Two additional requirements improve accuracy.  First, $\xi_{\rm crit} > 1$ in a small number of cases ($<1\%$), so we set a ceiling on $\xi_{\rm crit}$ so that the region $\xi > \xi_{\rm crit}$ contains at least five time slices and angle slices.  Second, we require that the region of integration be simply connected, so that it includes only the region around the central peak and no outlying autocorrelation peaks. These two requirements improve accuracy in test problems.  

\subsection{Tests}\label{sec:tests}

First, as a simple test, we use mock cylinder plots with known pattern speed (see also \citealp{Conroy_2023}).  We produced a set of 5000 realizations of a sheared, anisotropic Gaussian random field in $(\PA, t)$.  The shear sends $A(\PA,t) \rightarrow A(\PA - \Omega_m t,t)$, where $\Omega_m$ corresponds to the true pattern speed.  We then apply our technique to see if we can recover $\Omega_m$.  

Using our updated signal to noise threshold $\xi_{\rm crit}$ we are able to recover the correct pattern speed with a root mean squared error $\langle (\omegava - \Omega_m)^2/\Omega_m^2\rangle^{1/2} \sim 5\%$. For cylinder plots with a similar distribution of brightness temperature to the image domain near the critical curve, this is actually a slight reduction in accuracy (up from $\sim 3 \%$). However, our varying threshold is significantly more accurate when the distribution of features in the cylinder plot is quite different from the image domain, as is the case for small or large $\rho$ in the visibility domain (see Figure \ref{fig:cylinder}). This tradeoff is needed to generalize the pattern speed analysis to a range of $\rho$. 

$\omegava$ measurements become less accurate with increasing correlation angle $\PA_{\rm corr}$ (as the number of independent samples is reduced), decreasing correlation time $\tau_{\rm corr}$ (as features become poorly resolved in time), and increasing pattern speed (as the autocorrelation slope becomes poorly resolved in angle and time). Outliers with large error (percent error approaching or exceeding $100\%$) begin to occur when $\tau_{\rm corr} \lesssim 5 \,dt$ or $\PA_{\rm corr} \lesssim 5 \,d\PA$, for a cadence $dt$ and angular resolution $d\PA$. This motivates the $\xi_{\rm crit}$ ceiling that at least five time slices and angle slices be included in the integration bounds, as discussed in Section \ref{sec:method_pattern}. Nevertheless, for a typical $\PA_{\rm corr}$ and $\tau_{\rm corr}$, we can recover up to super-Keplerian values in $\Omega_m$.

\section{Results}\label{sec:results}

In the image domain, we found three results: (1) the pattern speed on the ring is sub-Keplerian (i.e., they are slower than a free body orbiting on a circular geodesic, and also slower than the plasma velocity which is subject to magnetic field and pressure gradient forces); (2) pattern speeds are dependent on inclination ($\Omega_p \propto \cos i$) and black hole mass ($\Omega_p \propto 1/M$); (3) the pattern speed on the ring is weakly dependent on spin; (4) pattern speed also depends on impact parameter, with declining pattern speed as one moves outward from the ring.

\subsection{$\uv$ Pseudo-rotation Curves} \label{sec:radial_dependence}

\begin{figure}
   \plotone{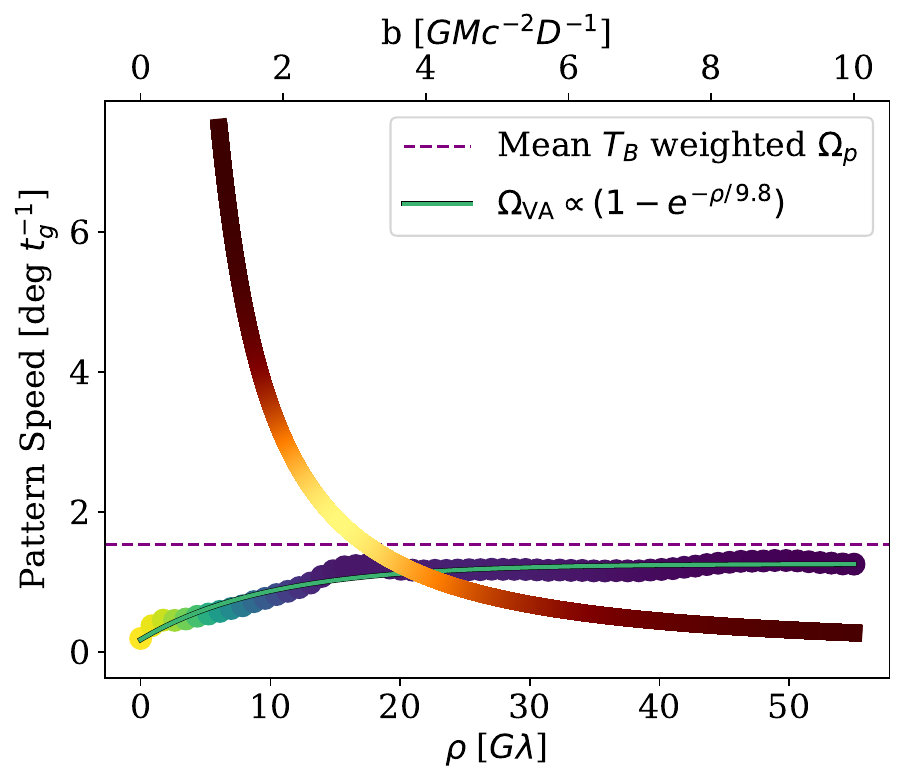}
   \caption{Pattern speed for a toy model in the image domain (shown in warm tones, with color denoting mean brightness at the angular radius, or impact parameter, marked by the top x-axis) and in the $\uv$ domain (cool tones, with color denoting mean amplitude for a given $\rho$ marked by the bottom x-axis).
   The image-domain values are measured from an analytic {\tt iNoisy} model with a known power-law dependence.  The best-fit curve to the $\uv$ pattern speed (blue) asymptotically approaches the brightness-weighted image-domain pattern speed (dashed purple). 
   }       
   \label{fig:iNoisy} 
\end{figure}

Here, we find that the sign of $\Omega_p$ is the same in the $\uv$ and image domain.  That is, clockwise image-domain motion is associated with clockwise $\uv$ motion, and likewise for counterclockwise motion.  This was expected based on the arguments presented in Section \ref{sec:expect}.

The Keplerian rotation frequency $\Omega_K$ at $r \simeq 4 \, GMc^{-2}$, where most of the ring emission is produced, is $\simeq 7$ deg $t_g^{-1}$, and  $\Omega_p < \Omega_K$.  We find a similar result in the $\uv$ domain, with $\omegava < \Omega_K(r = 4)$ on all baselines (at all $\rho)$.  

In the image domain, $|\Omega_p|$ increases with decreasing impact parameter \citep[see Figure 5 in][]{Conroy_2023}.   What do we expect for the $\rho$ dependence in the $\uv$ domain?  Figure \ref{fig:iNoisy} shows $\Omega_p(b)$ and $\omegava$ for a toy model with analytically known radial dependence: $\Omega_{p} \propto b^{-3/2}$. This model was made using the {\tt iNoisy} code, where the image is modeled as inhomogeneous, anisotropic Gaussian random fields  \citep{Lee_2021}.

The figure shows several features that generalize from the toy model to GRMHD models.  $\omegava$ is a minimum near $\rho = 0$ and rises toward a constant on long baselines; $\Omega_p(b)$ is a maximum at small impact parameter and falls toward zero at large impact parameter.  Evidently, the rotation signal increases past a few $G\lambda$.   

We find that $\omegava(\rho)$ is well fit by $W (1-e^{-\rho/a})$, where $W$ is a parameter-dependent constant.  In the toy ({\tt iNoisy}) model on long baselines, the fit asymptotes approximately to the brightness-weighted image-domain pattern speed $\overline{\Omega}_p$.  In Figure \ref{fig:iNoisy}, $\overline{\Omega}_p$ is marked as a dashed purple line, which crosses $\Omega_{p}(b)$ near the impact parameter of peak brightness (i.e., the peak of the ring).   For GRMHD models, $\uv$ trends are noisier than for {\tt iNoisy} models, although a small survey of GRMHD models suggests they also approach the brightness-weighted image-domain pattern speed on long baselines.

\subsection{$\uv$ Pattern Speed Parameter Dependence}

As in the image domain, we find that $\omegava$ depends mainly on inclination and mass. Weakly magnetized SANE models have a dependence on the electron temperature prescription, $\rhigh$. And both MAD and SANE models show a spin dependence in their pseudo-rotation curves $\omegava(\rho)$.  Here, we summarize the parameter dependence by fitting pseudo-rotation curves.  

Following Section \ref{sec:radial_dependence}, we fit $\omegava(\rho; i, \spin) = W(i,\spin, \rhigh) (1 - e^{-\rho/a})$. We find: \begin{equation} 
    \begin{aligned}
        &\omegava \simeq \cos i \,(0.3 - 0.3\spin \\
        & 
        + (0.5 + 0.5\spin)(1 - e^{-\rho/3.1 G\lambda}) ) 
    \end{aligned} \label{eq:MAD_bestfit} \quad \text{(MAD)}
\end{equation}
\begin{equation}
    \begin{aligned}
    \omegava \simeq  \cos i \,(0.7- 0.04\spin - 0.01 \rhigh \\ 
    + \, (0.8 + 0.4\spin)(1 - e^{-\rho/22.6 G\lambda}) )
    \end{aligned} \label{eq:SANE_bestfit}
    \quad \text{(SANE)}
\end{equation} 
in units of deg $t_g^{-1}$.

\begin{figure}
\plotone{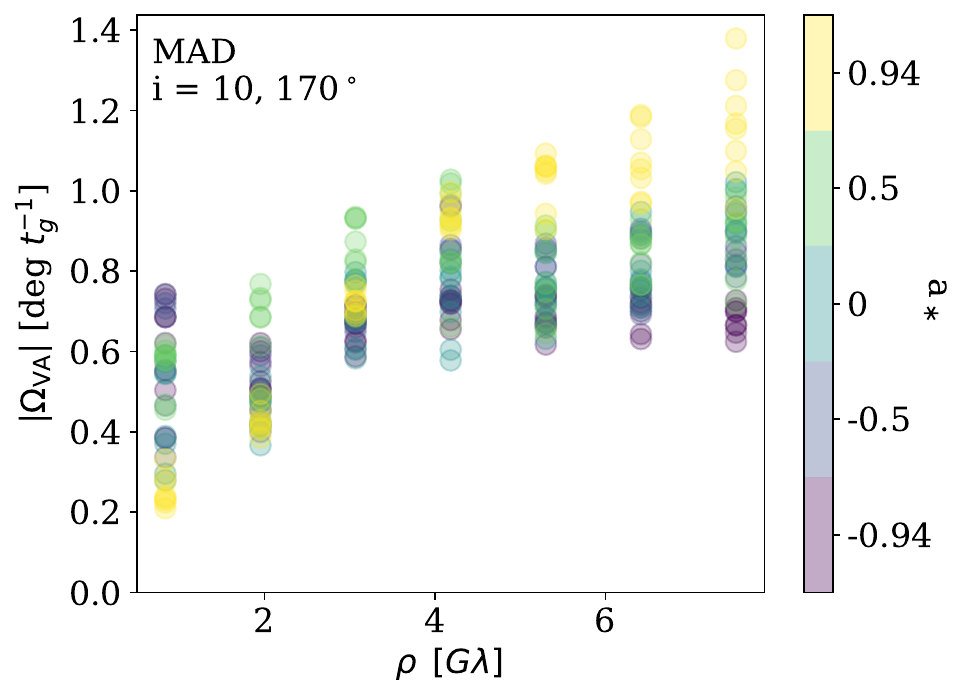}
   \caption{Measured $\uv$ pattern speeds as a function of baseline length, shown for all MAD face-on models ($i = 10^\circ, 170^\circ$). Color corresponds to dimensionless black hole spin $\spin$. Notice that prograde ($\spin > 0$) models slope more steeply, while retrograde ($\spin < 0$) models have a shallower slope.}       
\label{fig:mad_faceon} 
\end{figure}

\begin{figure*}
    \centering
    \plotone{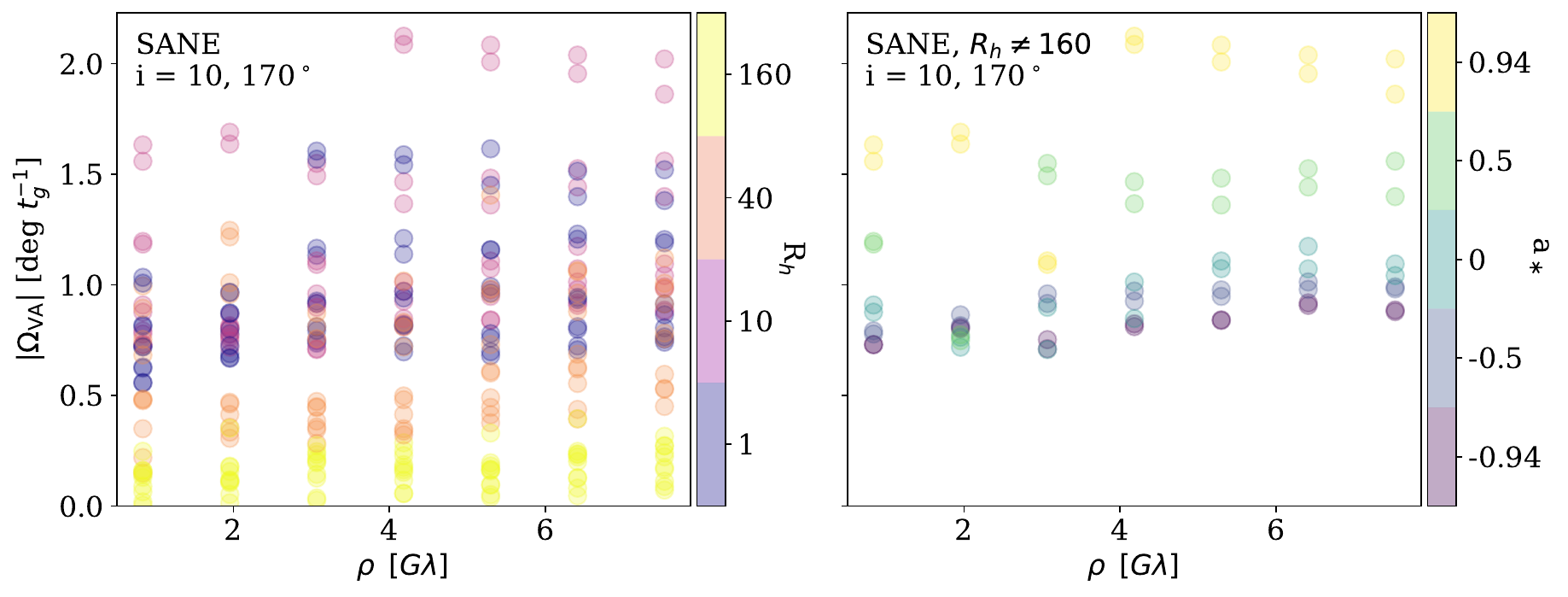}
    \caption{Measured $\uv$ pattern speeds as a function of baseline length, shown for all SANE face-on models ($i = 10^\circ, 170^\circ$). Color corresponds to the electron temperature distribution function (left) and black hole spin (right). To make the spin dependence clearer, we have not plotted $\rhigh = 160$ on the right plot. SANE models demonstrate a strong dependence on $\rhigh$ and a weaker dependence on $\spin$. 
        \label{fig:sane_faceon}
     }
\end{figure*}

As in the image domain, there is a strong dependence on inclination. The sign of $\omegava$ follows the sign of $\cos i$; as we go from $i = 0^\circ$ to $i = 180^\circ$, the apparent direction of rotation flips from counterclockwise to clockwise. At $i = 90^\circ$, the dominant direction of motion becomes horizontal. If the angular momentum vector is pointed up, then part of the foreground flow appears to move clockwise above the midplane and the other part counterclockwise below. 
The effects of lensing are symmetrical in the cylinder plot: background features are lensed both above and below the midplane, which projects as both counterclockwise and clockwise motion.
These effects cancel when averaging over $\PA$. Thus, there is no net apparent rotation for $i=90^\circ$. This carries over to this visibility domain:  $\omegava \propto \cos 90^\circ = 0$. 

The $\cos i$ dependence is simple but approximate. The true relation may be steeper than this near $i = 90^\circ$ and shallower near face-on \citep[see Figure 8 in][]{Conroy_2023}. The $\cos i$ dependence is surprising, since one might naively expect a hot spot on a circular orbit with period $P$ in flat space to appear to rotate at $\Omega_\mathrm{hs} = \mathrm{sign}(\cos i)\, 2\pi/P$. Even in this simple example, however, the apparent angular speed of the hot spot varies as one traces its motion on the sky, with lower speed as the spot is approaching or receding from the observer. Modern hot-spot models also exhibit $\Omega_p<\Omega_\mathrm{hs}$ for $i>0^\circ$ \citep[compare the slopes of the central autocorrelation peak in the left- and right-most panels of Figure 7 in][]{Zhu_2025}. In GRMHD models, multiple factors may contribute to the approximate $\cos i$ dependence.  
Brightness fluctuations arise from a spectrum of entropy and pressure fluctuations with varying coherence times. 
The inclination changes which parts of this spectrum are heavily weighted in the pattern speed signal.
This is because the viewing angle and lensing produce changes in 
(1) the relative brightness of fluctuations over $\PA$, 
(2) the angle and radius at which the line of sight pierces the disk, and
(3) the apparent angular speed of fluctuations.  
Interestingly, we see a weak signal for the latter effect in our synthetic data, though this may be difficult to detect observationally.   

There is also a dependence on mass, here shown in the units of $\omegava$ and Equations \ref{eq:MAD_bestfit}–\ref{eq:SANE_bestfit}. As the mass of the black hole increases, the dynamical time increases: it takes longer for waves to travel around a larger black hole. The pattern speed is thus inversely proportional to mass. When moving at a typical value of $\Omega_p \approx 1$ deg $t_g^{-1}$, a feature has a period of $\tau \approx 120$ min for Sgr A* and $\tau \approx 130$ days for M87*. 

In MAD models, we find two fitting constants that depend on spin. The first adjusts the y-intercept, which is inversely proportional to spin for face-on models. The second adjusts the slope. Strongly spinning prograde models ($\spin > 0$) have lower pattern speeds near $\rho \approx 0$, but have a sharply steeping rotation curve at small $\rho$. Strongly spinning retrograde models ($\spin < 0$) have a higher pattern speed near $\rho \approx 0$ that does not increase as much at higher $\rho$. These trends can be seen in Figure \ref{fig:mad_faceon}, which shows $\omegava$ in face-on MAD models. 

In SANE models, we likewise find a spin dependence in the slope and vertical shift of the $\uv$ pattern speed rotation curve. 
This can be seen in the right panel of Figure \ref{fig:sane_faceon}. 

SANE models exhibit strong dependence on the electron temperature distribution $\rhigh$; MAD models do not. This is because varying $\rhigh$ changes the latitude of emission in SANE models. There, the midplane is brightest at $\rhigh \approx 10$. The dominant emission region becomes slightly more diffuse and off-midplane for $\rhigh = 40$. Near $\rhigh = 160$, the emitting region shifts significantly toward higher latitudes and toward the jet-disk boundary \citep[see Figure 4 in][]{M87PaperV}. 

One can see the SANE dependence on $\rhigh$ in the left panel of Figure \ref{fig:sane_faceon}. It seems the apparent azimuthal motion near the funnel ($\rhigh = 160$) is slower than in the midplane. This makes sense. The accretion disk maintains a roughly constant angular velocity on spherical surfaces, and the angular velocity decreases with radius \citep{Wong_2021}. For a given impact parameter at  larger latitude, the source fluctuations lie at larger radius and are thus moving more slowly.   

The autocorrelation peak for $\rhigh = 160$ SANEs is thinner in $\Delta t$, suggesting fast features in the funnel may be shorter-lived. In this case, we may also not have the time resolution to pick out the faster features (our models have a cadence of $\Delta t = 5 t_g \approx 100$ s for Sgr A* and $45  \, \mathrm{hr}$ for M87*). There is significant shear at the jet-disk boundary, and plasma can flow from the disk into the jet \citep{Wong_2021}. This could cause features to shear away or change on-sky impact parameter more quickly than in the midplane, before we can resolve them in time within a cylinder plot. These dependencies on the emission region imprint themselves on the $\rhigh = 160$ SANE $\uv$ pattern speeds. In contrast, MAD models remain brightest in the midplane independent of $\rhigh$ \citep[see Figure 4 in][]{M87PaperV} and MADs do not show a strong $\rhigh$ dependence. 

\subsection{Comparison to the Image-Domain Pattern Speed} 

The image domain and $\uv$ domain pattern speeds are the same order of magnitude. At the critical curve, the pattern speed is significantly sub-Keplerian: $\Omega_{p} \approx \Omega_K(r = 4 \, GMc^{-2})/7 \approx 1$ deg $t_g^{-1}$. If EHT measures an $\omegava$ that is also significantly slower than $7$ deg $t_g^{-1}$, that is consistent with $\Omega_p$ also being sub-Keplerian. 

This analysis is limited, however, without constraining the impact parameter of the rotating features in the image domain. For example, a strongly sub-Keplerian spiral wave near the millimeter emission region at $r \sim 4 \, G M c^{-2}$
may have a similar pattern speed to a Keplerian feature (such as the hot spots described in Section \ref{sec:intro}) orbiting at a much larger radius. Dynamical imaging methods could distinguish 
these cases, but more work is needed to distinguish these cases with $\uv$ data alone. 

The existence of waves with sub-Keplerian pattern speeds may also be relevant to GRAVITY collaboration findings. GRAVITY fit a hot-spot model to Sgr A* centroid and polarization data during near-infrared flares,  
favoring Keplerian to mildly super-Keplerian motion \citep{GRAVITY_2018, GRAVITY_2020_flares, GRAVITY_2023}. While hot spots (and entropy fluctuations generally) may occur in the accretion flow, they may not be much brighter than their surroundings: synchrotron emissivity changes only slowly with temperature, and increased magnetic field strength reduces the electron number density due to pressure balance. GRMHD modeling suggests that hot spots such as flux tubes may be edge-brightened \citep{Porth_2021}. If the near-infrared centroid tracks the outer edge of the spot \citep[as in][]{Dexter_2020}, then GRAVITY data could be consistent with a more sub-Keplerian feature. This again motivates the extraction of the radius of features using $\uv$ data.

\begin{figure}
    \centering
   \plotone{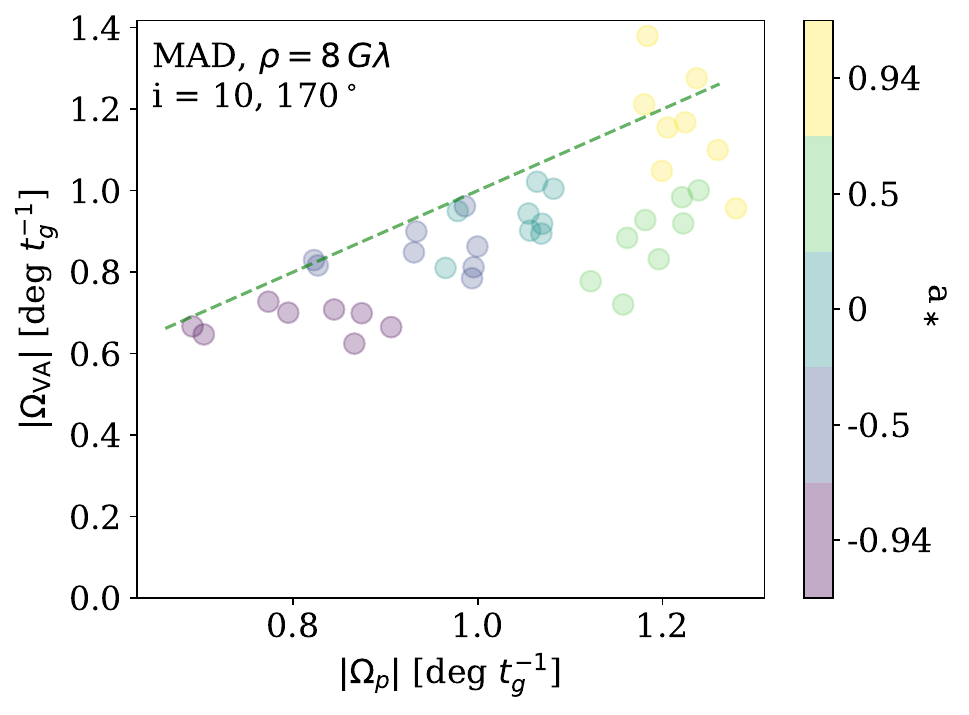}
   \caption{$\Omega_p$ at the critical curve vs $\omegava$ at $\rho \approx8 \, G\lambda$, for MAD face-on models ($i= 10, 170^\circ$). Color designates spin. The green dashed line marks $\Omega_{p}=\omegava$. 
   }       
   \label{fig:diff} 
\end{figure}

Given the similarity between $\Omega_{p}$ and $\omegava$, it is interesting to consider where $\uv$ pattern speed equals the image-domain pattern speed along the bright ring. In Figure \ref{fig:diff}, we plot $\Omega_{p}$ versus $\omegava(\rho \approx 8 \, G\lambda)$ for MAD face-on models. When $\spin = 0.94$, the steeper rotation curve means $\Omega_{p} \approx \omegava$ near $\rho = 8 \, G\lambda$. For retrograde models, the flatter rotation curve means $\Omega_{p} \approx \omegava$ farther out near $\rho \approx 25 \, G\lambda$.

\subsection{Connection to Features in the Image Domain} 
\label{sec:wavelet}

We might also ask what features in the image domain contribute to the visibility amplitudes to generate the $\uv$ pattern speeds. Current EHT coverage spans a range in baseline length: $\rho \in (0, \, \sim9) \, G\lambda$. Different points in $\uv$ space filter for different scales and $\PA$s in the image. We can map this dependence using wavelet convolution. 

\begin{figure*}
    \centering
   \plotone{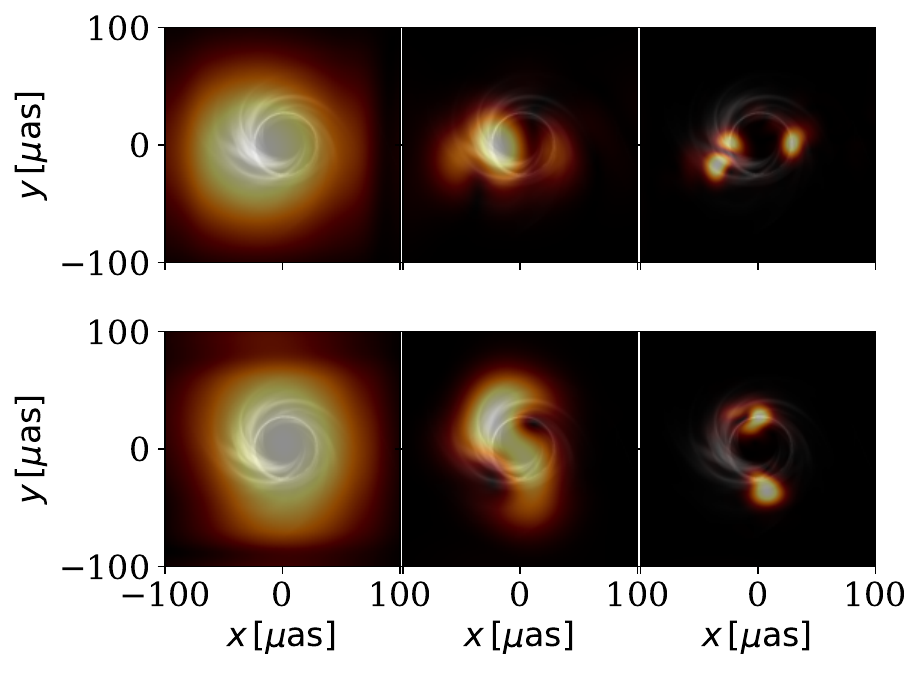}
   \caption{A frame from a typical GRMHD model (MAD, $\spin = 0.5$, $i = 30^\circ$, $\rhigh = 160$) convolved with a 2D Morlet wavelet, showing which image-domain features contribute to the visibility amplitude at $\rho = 4 \,G\lambda$ (left), $10 \,G\lambda$ (middle), and $30 \,G\lambda$ (right), filtering for horizontal (top) and vertical features (bottom). The original frame is shown in gray behind the convolution. As expected, large-scale features dominate the low-frequency $\uv$ data, while small-scale features dominate the high-frequency $\uv$ data.
   }       
   \label{fig:wavelet} 
\end{figure*}
Figure \ref{fig:wavelet} shows which regions of the underlying image dominate at various points in $\uv$ space (for $\rho = 4$, $10$, and $30 \, G\lambda$ and $\theta = 0$ and $90^\circ$). The figure was produced by convolving the underlying frame from Figure \ref{fig:frame} with a 2D Morlet wavelet, defined as
\begin{equation}
\phi(\vec{r}, \theta) = (1 / \sqrt{\pi} \lambda) \, e^{-i \vec{\omega_0} \cdot \vec{r}/\lambda} \, e^{-0.5|\vec{r}/\lambda|^2}
\end{equation}
Here $\vec{r}$ is a location vector, $\vec{\omega_0} = \langle \omega_0 \cos \theta,\omega_0 \sin \theta\rangle$ is a direction vector depending on the angle $\theta$ of your coverage in $\uv$ space, and $\lambda$ is a scale factor. Changes in $\omega_0$ shift whether we retain spatial or frequency information after convolution. 
Geometrically, it corresponds to the number of oscillations in the wavelet.
We set $|\omega_0| = 6$, a typical value, and
define $\lambda = (\omega_0 + \sqrt{2 + |\omega_0|^2})/(2 \rho)$.

At low $\rho$ (e.g. $\rho= 4 \, G\lambda$, as in the left panel of Figure \ref{fig:wavelet}), visibility amplitudes are driven by large-scale features in the image. These large-scale features change infrequently: the location of the ring does not change, and the brightness asymmetry $\PA$ is generally confined to a specific hemisphere \citep[see Figures 2, 3, and 5 in][]{M87PaperV}. Thus, it makes sense that the large-scale low-frequency pattern speed would have a lower magnitude. In the middle panel, just beyond the longest EHT baselines ($\rho = 10 \,G\lambda$), we find visibility amplitudes are dominated by a combination of the bright ring and groups of spiral features. Near $\rho = 30 \,G\lambda$, we find space–ground baselines would be sensitive to individual spiral waves. Since $\omegava$ increases with $\rho$, it seems individual spiral features propagate more quickly than groups of features or larger-scale features in the image.

\begin{figure*}
    \centering
    \plottwo{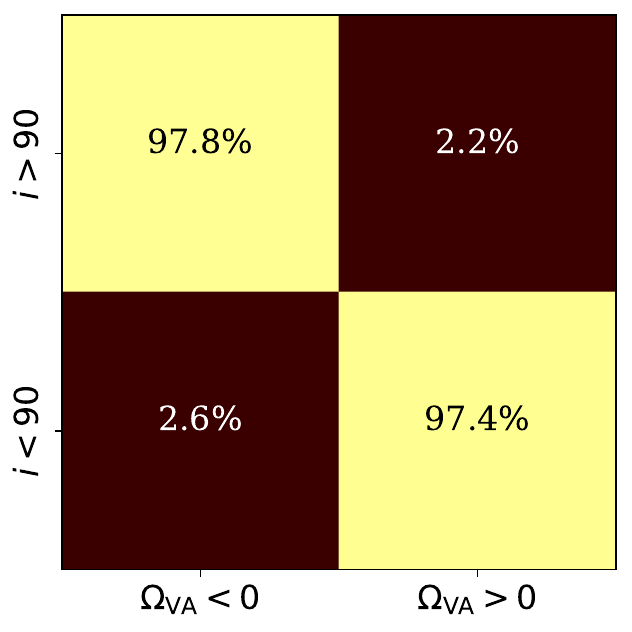}{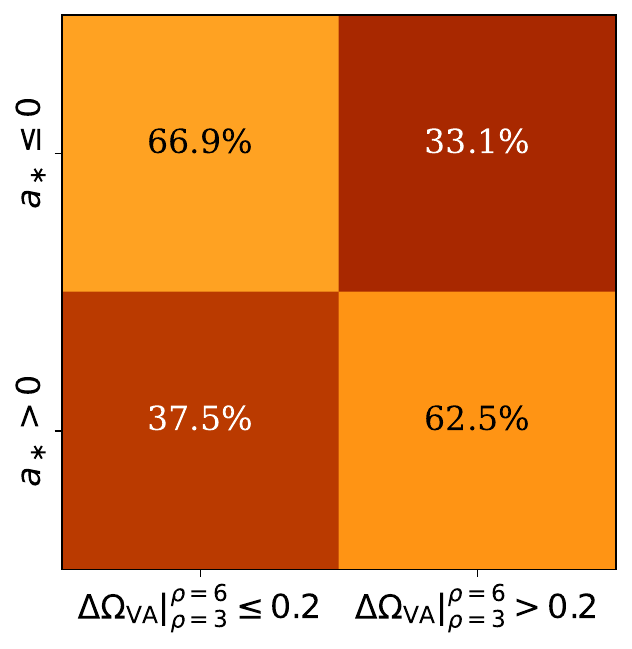}
   \caption{Constraints on the sign of $\cos \, i$ based on the direction of the $\uv$ pattern speed (left), and on the sign of $\spin$ based on how much the pattern speed magnitude increases from $\rho = 3$ to $6 \, G\lambda$ (right). Pattern speed measurements are from videos of MAD, non-edge on models ($i \neq 90^\circ$ left, $i \neq 70, 90,$ or $110^\circ$ right) with a $10 \, t_g$ cadence and a $300 \, t_g$ duration.
    }       
   \label{fig:spin_constraints_image_domain} 
\end{figure*}

In \citet{Conroy_2023}, we hypothesized that turbulence in the disk generates spiral waves. The rotation rate of the underlying plasma in the turbulent excitation region gets imprinted on the waves. If this interpretation is correct, then the larger pattern speed of smaller features would suggest these features are generated closer in, where the fluid velocity is larger. This would explain why $\omegava$  shows a greater spin dependence with increasing $\rho$, as the fluid velocity also becomes more dependent on spin closer in \citep[see, e.g., Figure 5 in][]{Conroy_2023}. There may be other interpretations consistent with the data. Regardless, this strong spin dependence at baselines far beyond current EHT coverage emphasizes the importance of space VLBI. 

\subsection{Spin Constraints}
\label{sec:spin_constraints}

Pattern speed measurements in the image and $\uv$ domain can provide spin constraints, but how precise are these constraints? Consider a movie with a cadence of $\Delta t = 10 \, t_g$ ($\approx 204$ s for Sgr A* and $\approx 3.7$ days for M87*) and a duration of $\Delta t = 300 \, t_g$ ($\approx 1.7$ hr for Sgr A* and $\approx 111$ days for M87*). This is a cadence and duration that can reasonably be achieved in future observations.

To assess spin constraints, we subdivide MAD GRMHD models with duration $\Delta t = 1.5 \times 10^4 \, t_g$ into subwindows with this more realistic observing cadence and duration and measure the $\uv$ pattern speed. 

The pattern speed is most sensitive to inclination. 
The simplest $\uv$ pattern speed measurement would be a detection of clockwise or counterclockwise motion, which would constrain the sign of $\cos \, i$ with high confidence.

The left panel of Figure \ref{fig:spin_constraints_image_domain} shows how often a measure of $\mathrm{sgn} \, \omegava(\rho = 6 \, G\lambda)$ over a $300 \, t_g$ observing window provides the correct sign of $\cos \, i$.  We measure the direction of rotation with near $100\%$ accuracy. The accuracy is lower on shorter baselines ($\sim 88\%$ at $\rho = 3 \, G\lambda$) and increases with baseline length ($\sim 94\%$ at $\rho = 4 \, G\lambda$). 
This accuracy when measuring $\mathrm{sgn} \cos i$ is slightly reduced in the updated ``v5" model library. 
Here, strongly retrograde models have more emission off the midplane, where fluid can rotate against the disk and with retrograde spin (overall, the rotation direction reverses $\sim 5\%$ more frequently across the v5 library; see Appendix \ref{subsec:v5} for discussion). 

Spin estimates would require multiple $\uv$ pattern speed measurements across multiple baselines to measure the pseudo-rotation curve. The right panel of Figure \ref{fig:spin_constraints_image_domain} shows what percentage of MAD, non-edge on models  increase in $\uv$ pattern speed by more than $0.2$ deg $t_g^{-1}$ from $\rho = 3$ to $6 \, G\lambda$ as a function of spin. This can constrain the sign of spin with $\sim 65\%$ accuracy. If we lower the threshold and simply ask how many models show any increase in $\uv$ pattern speed, we produce more accurate prograde constraints and less accurate retrograde ones ($81\%$ and $34\%$, respectively). Likewise, if we increase the threshold to $0.3$, we produce less accurate prograde constraints but more accurate retrograde ones ($53\%$ and $80\%$, respectively). 

To obtain higher-confidence spin constraints, one might incorporate complementary observable measurements. Possible observables include measurements of $\beta_2$ \citep{Palumbo_2020}, the inner shadow \citep{Chael_2021, Papoutsis_2023}, or the photon ring \citep{Johnson2020_photonring, Broderick_2022, Papoutsis_2023}. 

One especially interesting possibility is the ring's mean brightness asymmetry. The magnitude of the brightness asymmetry depends on the magnitude of the spin \citep{Bernshteyn_2026}, while the $\PA$ depends on the direction of spin on the sky. Surprisingly, the $\PA$ of the bright side of the ring follows the approaching side of the black hole spin, not the disk inclination \citep{M87PaperV, Conroy_2026_PA}. For a spinning black hole, GRMHD models predict
the mean brightness peak $\textrm{PA}$ of the ring to be $90^\circ$ counterclockwise from the position angle of the spin projected on the sky (i.e., on the approaching side of the black hole). In M87*, we have a high-confidence measurement of the mean brightness peak location \citep{M87PaperV, M87_2018_paperI}, which is consistent with GRMHD models if the spin vector is coaxial with the large-scale jet $\PA$ and inclination \citep{Walker_2018}. In Sgr A*, a detection of the mean brightness peak $\PA$ does not yet exist but may be possible in the future.

Thus, by measuring the direction of the spin on the sky ($\spin \cos \, i$) with the brightness asymmetry $\PA$, and by measuring the direction of the disk on the sky ($ \cos \, i$) with the pattern speed, it may be possible to constrain the sign of spin.  The $\uv$ pattern speed could thus complement other analyses, allowing for spin constraints without imaging.

\begin{figure*}
   \plotone{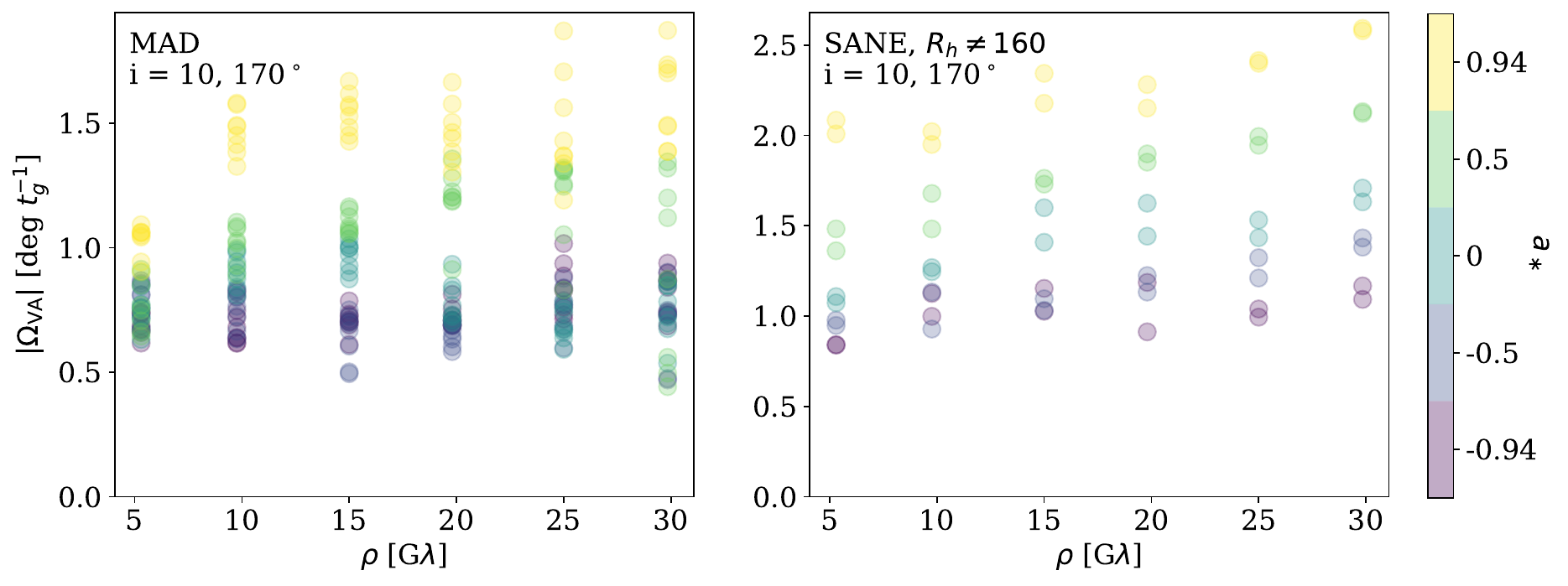}
   \caption{Measured $\uv$ pattern speeds as a function of baseline length $\rho$ for nearly face-on ($i = 10^\circ, 170^\circ$) MAD models (left) and SANE models (right), extended out past current EHT coverage. As in Figure \ref{fig:sane_faceon}, we filter out SANE $\rhigh = 160$, where $\omegava \approx 0$. EHT observations at $345 \, GHz$ would correspond to $\rho \lesssim 15 \, G\lambda$, while a space VLBI mission like BHEX would extend to $\rho \approx 30 \, G\lambda$. 
   }       
   \label{fig:Bhex_faceon} 
\end{figure*}

\section{Space VLBI}
\label{sec:space}

Efforts to fly a submillimeter VLBI antenna in space are now underway \citep{Bhex_22, BHEX_2024}.  This would provide access to long baselines with $\rho_{max} = 32.5 \, (a/(6.62 R_\oplus)) \, (\lambda/1.3 \,\mathrm{mm})^{-1} \, \mathrm{G}\lambda$ for a satellite with typical geocentric distance $a$, here normalized to the semimajor axis for geosynchronous orbits.  Since EHT sources have not been observed on these long baselines before, forecasting for space VLBI necessarily relies on models.  

For space VLBI, we propose a new method: a single-baseline $\uv$ pattern speed measurement. If we know the scale of $\uv$ features, we could measure the time it takes for features to pass over our track and use this to measure the mean rotation.

There are three timescales relevant to observations of changes in the feature on long baselines: 
(1) the time required for the satellite orbit to move across a coherent patch, 
\begin{equation}
\begin{split}
\sim \tau_{orb} \equiv \frac{\Delta\rho}{\Omega \rho} \qquad 
\qquad \qquad \\
= 680 \left( \frac{\Delta \rho}{1.6 \times 10^9} \right) \left( \frac{a}{6.62 R_\oplus} \right) ^{1/2} \frac{\lambda}{1.3 \, \mathrm{mm}} \, \mathrm{sec}
\end{split}
\end{equation} 
(2) the correlation time for the features (defined as the correlation time following the motion of the features), $\tau_{coh}$; and (3) the time required for rotation of the source to move a feature across a point in the UV domain, 
\begin{equation}
\begin{split}
\sim \tau_{rot} &\equiv \frac{\Delta\rho}{\Omega_p \, \rho} \approx 
 56 \, \frac{\Omega_p \, t_g}{\mathrm{deg}} \\
& \times \frac{M}{4 \times 10^6 M_\sun} 
\frac{\lambda}{1.3 \, \mathrm{mm}} 
\frac{\Delta \rho}{1.6 \times 10^9} \frac{a}{6.62 R_\oplus} \,{\rm sec}
\end{split}
\end{equation}
The coherence time for the patches is tens of $t_g$ in our models. For Sgr A*, the source will change very quickly for geostationary orbits due to rotation, and fringes must be monitored on a timescale of tens of minutes, which is practical.     

For M87*,  $\tau_{orb}$ is nearly the same, since the source has a similar size (so the coherent patches in visibility amplitude are of similar size) and the orbit is the same.  However, the correlation time is $\sim 1500$ times longer, as is $\tau_{rot}$, meaning that the satellite is dragging its $\uv$ probe across the pattern.  The motion of the satellite dominates, and it is not possible to measure the rotation speed from a single orbit.  Instead, it will be possible to cross-correlate fluctuations in visibility amplitudes between orbits.  

On long baselines that are inaccessible to EHT, the signal to noise decreases but the effect of spin on the $\uv$ pattern speed increases. Current EHT coverage at $230 \, GHz$ observations covers $\rho \approx 9 \, G\lambda$. At $345 \, GHz$ this would extend to $\rho \lesssim 13 \, G\lambda$, and a space VLBI mission such as the Black Hole Explorer (BHEX) could extend coverage to $\rho \approx 30 \, G\lambda$ \citep{Bhex_22, BHEX_2024}. 
On these longest baselines, visibility amplitudes are primarily determined by the photon ring and by individual spiral features. The motion of these features has a clear spin dependence in our models. Thus, a measurement of the $\uv$ pattern speed on long baselines (and the pseudo-rotation curve with respect to shorter baselines) could improve spin constraints. In Figure \ref{fig:Bhex_faceon}, we illustrate this by extending the pseudo-rotation curve for both MAD and SANE face-on models to $\rho = 30 \, G\lambda$.

\section{Conclusions}\label{sec:conclusion}

This paper extends our analysis of apparent rotation, or pattern speeds, in GRMHD models of EHT sources to the $\uv$ domain.  We measure the visibility amplitude at baseline length $\rho$ as a function of $\PA$. Using an autocorrelation method, we measure a mean apparent rotation rate at each $\rho$. Compared to image-domain analysis \citep{Conroy_2023}, we have also updated our methodology using a new technique to more optimally extract the pattern speed from noisy data. 

For all models in the Illinois Sgr A* image library, we find the visibility amplitude pattern speed $\omegava$ to be slower than the Keplerian velocity at the radius where most of the characteristic ring of emission is produced. The order of magnitude of $\omegava$ is consistent with the sub-Keplerian $\Omega_p$ in the image domain, although it varies strongly with baseline length.  $\omegava$ increases with $\rho$. Its sign and magnitude depends on the viewing inclination.  $\omegava$ also depends on mass in all models.  Weakly magnetized (SANE) models have a dependence on the electron temperature parameter $\rhigh$. We find that spin affects the structure of the pseudo-rotation curve $\omegava(\rho)$; this dependence is driven by changes in which image-domain features dominate as $\rho$ changes. The parameter dependencies are summarized in Equations \ref{eq:MAD_bestfit} and \ref{eq:SANE_bestfit}.

In M87*, a $\uv$ pattern speed measurement would produce a distance-independent constraint on mass, assuming a known inclination from the angle of the large-scale jet. Since we know the sign of $a_* \cos i$ based on the mean $\PA$ of the brightness peak 
\citep[see Figure 5 in][]{M87PaperV},
a measurement of the direction of motion would measure $\textrm{sgn}(\cos i)$, thus revealing $\textrm{sgn} (a_*)$ (i.e., whether M87* is prograde or retrograde).

In Sgr A*, a $\uv$ pattern speed measurement would provide a constraint on inclination, given the strong prior on mass from stellar orbit measurements. A measurement of the direction of motion would allow us to test the connection between fluctuations in the millimeter-wavelength source and the near-infrared source, which appears to be rotating clockwise in measurements of the motion of the centroid of emission by the GRAVITY collaboration \citep{GRAVITY_2018, GRAVITY_2020_flares}.

In this paper, we assumed complete $\uv$ coverage to map out trends in $\uv$ pattern speeds with model parameters and baseline length. This will not be achieved in actual EHT data. Incomplete $\uv$ and time coverage creates gaps. 

In M87*, the dynamical time is much longer than one Earth rotation ($\tau \approx 50 \, t_g \approx 135$ days). So, measured visibility amplitudes seen along each nightly track are nearly time slices in cylinder plots (see Figure \ref{fig:cylinder}).  To measure azimuthal motion in the $\uv$ plane, we will need to correlate the same tracks on subsequent days.

For Sgr A*, the dynamical time is much shorter than one Earth rotation ($\tau \approx 17$ minutes).  Then, measured visibility amplitudes for each track are  nearly horizontal $\PA$ slices in cylinder plots.  Thus, $\uv$ pattern speed measurements require correlating adjacent baselines. Well-chosen baselines  should have minimal separation in $\rho$, no larger than the coherence scale of visibility amplitudes: half-width $\Delta\rho \lesssim (2\pi\sigma)^{-1}\approx 1.6 \, \textrm{G}\lambda$ (see Section \ref{sec:space}). The baselines should also be close enough in $\PA$ that features remain correlated as they cross each sampling point. Changes in $\rho$ over time may introduce bias.

In both sources, $\uv$ pattern speed measurements are possible by correlating two tracks via lagged interferometric covariance. Care should be taken when selecting tracks. The correlation peaks may require corrections from bias (e.g. from the changing total image flux, emission structure, observing parameters, etc.). \citet{Johnson_2015} described this method in greater detail. 

These results are limited by model uncertainty. Our earlier work finds that limited cadence and the fast light approximation used in ray-tracing our model images introduces only small errors \citep{Conroy_2023}.
Our findings are consistent: the same  GRMHD model exhibits $\omegava(\rho=5\,G\lambda)=0.11$ deg $t_g^{-1}$ under the fast light approximation and $0.10$ deg $t_g^{-1}$ with the more accurate slow light treatment.

There are potentially other sources of systematic error. In this paper, we considered only ideal GRMHD simulations (neglecting nonideal effects such as viscosity, heat conduction, and resistivity, as well as modified theories of gravity). We might also consider other parameters that we have not varied here. These include black hole tilt, additional electron distribution function parameters, GRMHD resolution (which might resolve smaller-scale features), and boundary conditions \citep[e.g.][]{ressler_2020}.  However, an initial inspection of nonideal simulations, model libraries with additional parameters, and models with differing resolutions and boundary conditions yields similar conclusions: pattern speeds are still sub-Keplerian; the pattern speeds still show a strong dependence on inclination and mass; and pattern speeds still show a weak dependence on spin through the pattern speed rotation curve. 

\section{Acknowledgements} \label{sec:Acknowledgements}
This work was supported by the NSF grant AST 20-34306.  This research used resources of the Oak Ridge Leadership Computing Facility at the Oak Ridge National Laboratory, which is supported by the Office of Science of the U.S. Department of Energy under Contract No. DE-AC05-00OR22725.  This research used resources of the Argonne Leadership Computing Facility, which is a DOE Office of Science User Facility supported under Contract DE-AC02-06CH11357.  This research was done using services provided by the OSG Consortium, which is supported by the National Science Foundation awards \#2030508 and \#1836650.  This research is part of the Delta research computing project, which is supported by the National Science Foundation (award OCI 2005572), and the State of Illinois. Delta is a joint effort of the University of Illinois at Urbana-Champaign and its National Center for Supercomputing Applications.

N.C. is supported by the NASA Future Investigators in NASA Earth and Space Science and Technology (FINESST) program. This material is based upon work supported by the National Aeronautics and Space Administration under Grant No. 80NSSC24K1475 issued through the Science Mission Directorate.

We thank Maciek Wielgus and the anonymous referee for providing comments that greatly improved this paper. We thank George Wong for providing the slow light model used in Section \ref{sec:conclusion}.

\appendix

\section{Updates to the Sgr A* Library}
\label{subsec:v5}

The main text analyzes data from the ``v3'' version of the Illinois Sgr A* library, which was also used in \citet{SgrAPaperI} and \citet{Conroy_2023}.\footnote{The $\omegava$ data in this work is available on the Illinois Data Bank: DOI \href{https://doi.org/10.13012/B2IDB-7205303_V1}{10.13012/B2IDB-7205303\_V1} \citep{Conroy_2025rawdata}.} Since then, we have produced a ``v5" library, which includes a dense spin sampling ($\spin = \pm 0.97, \pm 0.94, \pm 0.85, \pm 0.75, \pm 0.5, \pm 0.25,$ and $0$), disk tilt where the disk angular momentum vector does not align with the spin axis ($\phi =$ 0$^\circ$, 10$^\circ$, 20$^\circ$, and 30$^\circ$), and an updated adiabatic index ($\gamma = 4/3 \rightarrow 5/3$).

The v5 library was not used for this work, as it has not yet been densely sampled in inclination (only $i =$ 30$^\circ$, 50$^\circ$, and 90$^\circ$). In an initial survey, we found  qualitatively similar parameter dependencies: $\Omega_p$ and $\omegava$ depend on inclination ($\propto \cos i$), and are inversely proportional to mass, and there is still a spin dependence in the pseudo-rotation curve.

However, we do find a systematic decrease in the pattern speed magnitude in the v5 library. In prograde models, we find a reduction in the pattern speed by a factor of $\sim 2$. Consider our fiducial v3 model (MAD, $\spin = 0.5$, $i = 30^\circ$, $\rhigh = 160$). Both the observed image-domain pattern speed and the pattern speed of midplane pressure fluctuations ($r = 3 \, GMc^{-2}$) are $1.1$ deg $t_g^{-1}$ \citep{Conroy_2023}. In one $5,000 \,t_g$ interval of the same model in the v5 library, the image-domain pattern speed is $\Omega_p = 0.57$ ($\approx 0.66$ when projected to $i=0^\circ$). In retrograde models, we see a greater decrease in the magnitude of the pattern speed.  The change in pattern speed may be driven by two effects.

First, the increase in adiabatic index leads to higher temperatures in the emission region near $r \sim 4 G Mc^{-2}$, more pressure support of the flow, and thus slower rotation.   

Second, the slower pattern speed in v5  may be driven by an increase in off-midplane emission. 
The angular fluid velocity tends to decrease roughly as a function of radius \citep[e.g.,][]{Wong_2021}. For a given impact parameter on the sky, the increase in off-midplane emission causes us to be more sensitive to features at a larger radius, which are thus moving more slowly.

When this off-midplane emission extends into the jet boundary in retrograde models, we introduce a new phenomenon: counterrotation. Since a Blandford–Znajek jet is powered by the black hole spin and not the disk \citep{Blandford_Znajek}, our retrograde models produce jets that rotate with the black hole spin and not the disk \citep[Conroy et al. in preparation]{Wong_2021}. In v5, counterrotating features in the jet are visible. These get superimposed on corotating features in the disk, causing the mean magnitude of the pattern speed to decrease. 

\begin{figure}
    \centering
   \plotone{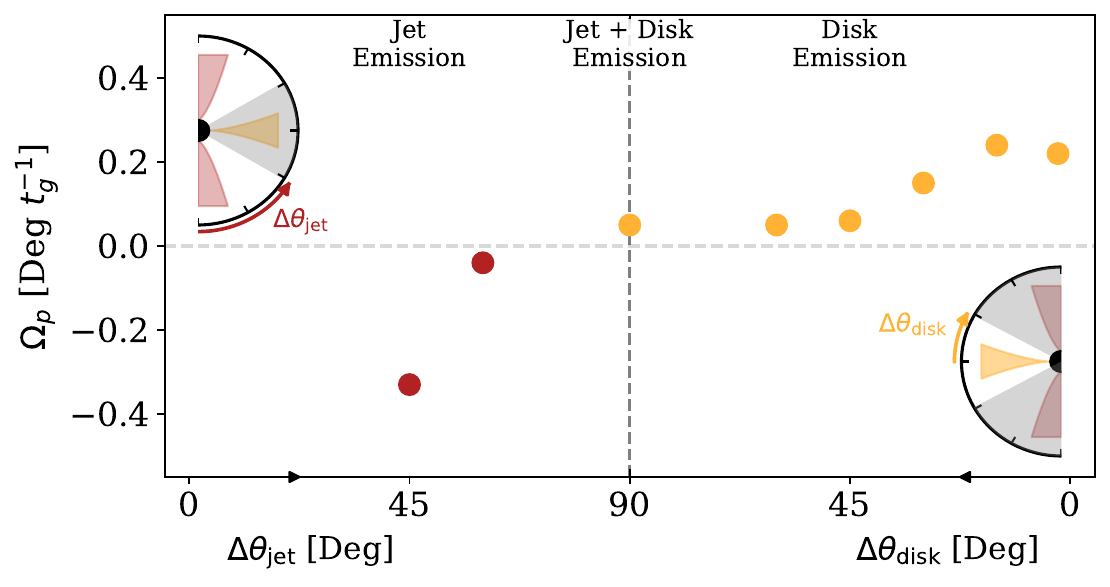}
   \caption{$\Omega_p$ in a v5 retrograde model (MAD, $\spin = -0.5$, $i = 30^\circ$, $\rhigh = 160$)
   as a function of different emission masks. Left: we only include emission within $\Delta \theta_{\mathrm{jet}}$ of the $z$ (jet) axis. Right: we only include emission within $\Delta \theta_{\mathrm{disk}}$ of the midplane. When $\Delta \theta_{\mathrm{jet}}=\Delta \theta_{\mathrm{disk}}=90^\circ$, we plot the mean $\Omega_p$ from all emission. 
   Black hole diagrams in the top-left and bottom-right corners illustrate $\Delta \theta_{\mathrm{jet}}$ and $\Delta \theta_{\mathrm{disk}}$.  
   As $\Delta \theta_{\mathrm{jet}}$  decreases, we filter for jet emission only and thus see counterrotation ($\Omega_p < 0$ for $i=30^\circ$).  Jet counterrotation is a fundamental prediction of the Blandford–Znajek jet-launching mechanism when $\spin<0$.
   }       
   \label{fig:BZ_jet} 
\end{figure}

To demonstrate this effect, consider a test model: MAD, $\spin = -0.5$, $i = 30^\circ$, $\rhigh = 160$. {\tt ipole} allows us to filter for emission just within a $\theta$ range around the midplane. We re-imaged our test model with masks of half-width $\Delta \theta_{\mathrm{disk}} \in \{2.5^\circ, 15^\circ, 30^\circ, 45^\circ, 60^\circ, 90^\circ \}$ centered around the midplane (where $\Delta \theta_{\mathrm{disk}} = 90^\circ$ corresponds to emission from all values $\theta$)
for a duration of $\Delta t = 1000 \, t_g$. We see $\Omega_p = \{0.22, 0.24,  0.15, 0.06, 0.05, 0.05\}$ deg $t_g^{-1}$ respectively. Evidently, as we include more off-midplane and jet emission, the measured pattern speed decreases. The pattern speed of midplane pressure fluctuations still corresponds well to the midplane pattern speed seen in images. Conversely, we can also test whether features in the jet have a negative pattern speed, as predicted. 
For features within $\Delta \theta_{\mathrm{jet}} = 60$ and $45^\circ$ of the $z$ (jet) axis, we observe $\Omega_p = -0.04$ and $-0.33$ deg $t_g^{-1}$ respectively. The more we filter for retrograde jet features, the more negative the pattern speed becomes. These counterrotating features explain why retrograde v5 models exhibit a greater decrease in $|\Omega_p|$ than prograde v5. As an aside, note that counterrotation in retrograde jets may act as a prediction from Blandford–Znajek, and thus may be used as a test of Blandford–Znajek. 
These results are shown in Figure \ref{fig:BZ_jet}.

Overall, uncertainties in the adiabatic index and jet emissivity contribute to systematic uncertainties in the pattern speed magnitude. This motivates future research \citep[see, e.g.,][for a study of Sgr A*'s adiabatic index]{Gammie_2025}. Still, both v3 and v5 libraries predict the EHT will measure sub-Keplerian pattern speeds in black hole videos. Pattern speeds in both libraries depend on inclination, mass, and spin, suggesting we could constrain physical parameters using the pattern speed, in conjunction with other observables. Finally, off-midplane emissivity in v5 suggests the EHT could detect counterrotating features in retrograde models, providing a novel route for measuring black hole spin and testing Blandford-Znajek.  

\FloatBarrier
\bibliography{main}{}
\bibliographystyle{aasjournal}

\end{document}